\documentclass[a4paper,twoside,twocolumn,english,american,superscriptaddress]{revtex4}
\usepackage[T1]{fontenc}
\usepackage[latin9]{inputenc}
\usepackage{amstext}
\usepackage{amssymb}
\usepackage{graphicx}

\makeatletter

\providecommand{\tabularnewline}{\\}
\newcommand{\lyxdot}{.}

\@ifundefined{textcolor}{}
{%
 \definecolor{BLACK}{gray}{0}
 \definecolor{WHITE}{gray}{1}
 \definecolor{RED}{rgb}{1,0,0}
 \definecolor{GREEN}{rgb}{0,1,0}
 \definecolor{BLUE}{rgb}{0,0,1}
 \definecolor{CYAN}{cmyk}{1,0,0,0}
 \definecolor{MAGENTA}{cmyk}{0,1,0,0}
 \definecolor{YELLOW}{cmyk}{0,0,1,0}
}

\@ifundefined{showcaptionsetup}{}{%
 \PassOptionsToPackage{caption=false}{subfig}}
\usepackage{subfig}
\makeatother

\usepackage{babel}
\begin{document}

\title{Beyond two-stage models for lung carcinogenesis in the Mayak workers:
Implications for Plutonium risk}

\author{Sascha Zöllner}

\email{sascha.zoellner@helmholtz-muenchen.de}

\selectlanguage{american}%

\affiliation{Helmholtz Zentrum München, Institute of Radiation Protection, Neuherberg
(Germany)}

\author{Mikhail E. Sokolnikov}

\affiliation{Southern Urals Biophysics Institute, Ozyorsk (Russia)}

\author{Markus Eidemüller}

\affiliation{Helmholtz Zentrum München, Institute of Radiation Protection, Neuherberg
(Germany)}

\date{\today}
\begin{abstract}
Mechanistic multi-stage models are used to analyze lung-cancer mortality
after Plutonium exposure in the Mayak-workers cohort, with follow-up
until 2008. Besides the established two-stage model with clonal expansion,
models with three mutation stages as well as a model with two distinct
pathways to cancer are studied. The results suggest that three-stage
models offer an improved description of the data. The best-fitting
models point to a mechanism where radiation increases the rate of
clonal expansion. This is interpreted in terms of changes in cell-cycle
control mediated by bystander signaling or repopulation following
cell killing. No statistical evidence for a two-pathway model is found.
To elucidate the implications of the different models for radiation
risk, several exposure scenarios are studied. Models with a radiation
effect at an early stage show a delayed response and a pronounced
drop-off with older ages at exposure. Moreover, the dose-response
relationship is strongly nonlinear for all three-stage models, revealing
a marked increase above a critical dose.
\end{abstract}
\maketitle

\section{Introduction}

Cancer is a genetic disease. In the widely held theory of somatic
evolution \cite{weinberg2013biology}, a cell's path toward the malignant
state is portrayed as a series of mutations or epigenetic events,
lending it successive selective advantages. These advantages, as summarized
in the ``hallmarks of cancer'' \cite{Hanahan2011}, essentially
amount to an increasingly uncontrolled proliferation. 

Those essential features---mutations accompanied by proliferation---have
long been identified as key ingredients in modeling carcinogenesis.
Beginning with the seminal multi-step models by Armitage/Doll and
Nordling \cite{nordling1953new,armitage1954age}, this eventually
led to the stochastic two-stage model with clonal expansion due to
Moolgavkar, Venzon, and Knudson \cite{Knudson1971,Moolgavkar1979},
which has by now become an established tool to understand and predict
cancer risk \cite{tan2008handbook,Little2010a,Jacob2010}. 

What fundamentally distinguishes such mechanistic models from conventional
epidemiological ones is that they do not directly model the endpoint---say,
the cancer mortality rate---but rather the process thought to lead
to it, parametrized via mutation and proliferation rates. This may
prove useful especially when the mortality rate is highly convoluted
by an exposure to carcinogens such as ionizing radiation, as is the
case in this study.  

Such a mechanistic approach can be readily generalized so as to build
in known biological effects, such as multiple genetic pathways \cite{Little2003,tan2008handbook}
or a more realistic number of stages \cite{Little1995,Luebeck2002}.
Indeed, for colorectal cancer, where the understanding of the cellular
mechanisms is comparatively advanced \cite{lengauer1997genetic,Nowak2002},
a number of extended models have been put forward to account for the
role of genomic instability \cite{Little2003,Little2007,Little2008,kaiser2014genomic},
the rather large number of premalignant stages \cite{Luebeck2002,Meza2008},
as well as the intricate dynamics during progression \cite{luebeck2013impact}.

By contrast, far less is known regarding other cancer types. For lung
cancer, mechanistic modeling studies are abundant but have focused
almost exclusively on the two-stage model. These indicate that for
the two main risk factors, smoking \cite{Hazelton2005,Schollnberger2006,meza2008analysis}
and $\alpha$-particle radiation (most relevant being Radon decay
products) \cite{Luebeck1999,Leenhouts1999,hazelton2001analysis,Brugmans2004,Heidenreich2004,VanDillen2011,Heidenreich2012,Eidemuller2012,Jacob2005,Jacob2007},
the best description is afforded by an enhancement of the proliferation
rate of premalignant cells. However, for radiation, this conclusion
has been disputed \cite{brugmans2002overrated} because it lacks a
conclusive biological mechanism, in contrast to the accepted mutagenic
effect of radiation. This debate has been further sparked by the single
analysis to date going beyond the two-stage model \cite{Little2002}.
Comparing fits to the Colorado-miners data using the two-stage model
with those from a subclass of three-stage models, the authors suggested
that a proliferation effect were confined to the two-stage model,
whereas a better fit quality was achieved within a three-stage framework
with a mutational radiation action.

The objective of this paper is to perform a comparative analysis of
lung-cancer risk associated with $\alpha$-radiation using different
multi-stage models -- specifically two- and three-stage models as
well as a two-path model for (radiation-induced) genomic instability.
Our goal is to identify the mechanisms of radiation action suggested
by those models, as well as to lay out to what extent their predicted
radiation risks are consistent. To this end, we apply these models
to the Mayak-workers cohort \cite{koshurnikova1999characteristics,Anspaugh2002}.
These workers, employed at the formerly Soviet Plutonium-production
plant, have been exposed to substantial doses of $^{239}\mathrm{Pu}$
via inhalation and exhibit a large number of lung-cancer deaths, 895
in total \cite{Gilbert2013}. A notable feature of Plutonium exposure
is its strong protraction, which might facilitate the assessment of
risk on the long time scales relevant to indoor Radon. Furthermore,
information on the strongest risk factor, smoking, is available for
most workers. We will show that certain three-stage models give an
improved description of the data, and we elucidate how these lead
to predictions for the risk that are qualitatively different from
both two-stage mechanistic and standard descriptive models.

\emph{}

\section{Materials and Methods}

\subsection{Mayak-workers cohort}

The Mayak-workers cohort comprises nuclear workers at the Mayak Plutonium-production
facility at Ozyorsk, Russia \cite{koshurnikova1999characteristics}.
The current follow-up includes all years 1948--2008 and comprises
25,757 members, cf. Ref.~\cite{Gilbert2013} for a comprehensive
overview. 

Many of the workers have been exposed to Plutonium-239, predominantly
through inhalation. These internal doses have been assessed via urine
measurements combined with biokinetic modeling for about 40\% of workers
in the plants at risk \cite{Khokhryakov2013}. Exposure to external
gamma radiation has been recorded via film-badge dosimeters \cite{Vasilenko2005},
and average annual dose rates are available for all workers. Furthermore,
for most workers, information exists on smoking status (non-/eversmoker)
as well as on alcohol consumption (teetotaler/light/medium/heavy/chronic),
see below. Owing to pronounced smoking and Plutonium-inhalation patterns,
the dominant cancer-mortality endpoint is lung cancer (defined here
as ICD-9 code 162), with a total of 895 mortality cases.

\subsubsection{Cohort definition }

To obtain a sufficiently homogeneous data set amenable to mechanistic
modeling, several selection criteria have been applied, similar to
previous studies \cite{Jacob2005,Jacob2007}. Our reduced (sub\-)cohort
excludes females, since these make up less than 25\% of the whole
cohort and exhibit very different mortality rates. Moreover, full
information is required on smoking/alcohol status and annual internal
dose(rate)s -- i.e., $^{239}\mathrm{Pu}$ doses must be measured or
assumed to vanish (for workers outside the radiochemical and Pu plants).%
\footnote{Missing risk-factor information may be taken into account via additional
categories. However, we found that this does not reduced the fit errors
due to the noise introduced this way.%
} 

Finally, the follow-up period is restricted as follows. If a Pu measurement
has been performed at time $t_{\mathrm{Pu}}$, then the entry date
is set to $t_{\mathrm{Pu}}+2\mathrm{a}$. This is done to avoid selection
bias, specifically healthy-survivor effects (due to extended follow-up
periods for persons surviving until $t_{\mathrm{Pu}}$) and diagnostic
bias (in case the measurement has been caused by imminent health problems).
To ensure complete follow-up, the exit date is cut off at the end
of 2008 or, in the case of migrants, 2003.

The reduced cohort includes 8,604 persons and 388 lung-cancer deaths.

\subsubsection{Risk factors}

Let us briefly highlight some aspects of the major risk factors (see
Ref.~\cite{Gilbert2013} for details). The main interest here is
in internal radiation, with measured nonzero doses available for 3,667
persons. Due to slow degradation in the lungs and the long half-life
of $^{239}\mathrm{Pu}$, exposures are highly protracted: First exposure
peaks around age 20, and typically continues until the end of follow-up.
Cumulative lung doses are well described by a log-normal distribution.
Among those measured, it is peaked at 4mGy, much below the mean dose
$\bar{D}=0.12\mathrm{Gy}$. Restricted to lung-cancer cases, the dose
distribution is shifted to higher values, with $\bar{D}=0.44\mathrm{Gy}$
and lower/upper 5\% quantiles of 7.6mGy and 2.3Gy. 

The overall smoking fraction is about $3/4$. Alcohol status, although
not known to be a risk factor for lung cancer, may serve as an indicator
for smoking habits because the fraction of smokers increases with
alcohol consumption. We group heavy/chronic drinkers as one category,
$a=1$, and otherwise set $a=0$.

As with any cancer, age is a crucial intrinsic risk factor. The ages
of cohort members range broadly between about 18 and 81 years, as
defined by the lower/upper 5\% quantiles of entry/exit age, with an
average of $27$ years spent in the cohort. By contrast, 90\% of lung-cancer
cases are found only between ages $49-78$, with a mean cancer age
of 65 years.

\subsubsection{Ethics statement}

The study of the Mayak-workers cohort has been reviewed and approved
by the Southern Urals Biophysics Institute's Review Board for issues
related to privacy and personal data protection.

\subsection{Statistical analysis}

\subsubsection{Multi-stage models}

Different multi-stage models are applied to the data. Their common
rationale is to radically reduce the complexity of carcinogenesis
to essentially two key processes: (i) \emph{mutations}---or, generally,
a series of (epi-)genetic transitions from healthy via pre-malignant
to malignant stem-like cells--- and (ii) \emph{proliferation }(i.e.,
symmetric division; cell inactivation or death) of pre-malignant cells
with a selective advantage \cite{Moolgavkar1999}.%
\footnote{This simplified single-cell picture does not explicitly include cellular
interactions. In particular, non-stem cells are not taken into account.
Thus, space limitations are not modeled, and \emph{cell }growth is
exponential rather than logistic. By contrast the \emph{hazard }is
bounded as an increased number of premalignant cells leads to more
cancer cases, hence reducing the population at risk.%
} 

\selectlanguage{english}%
\begin{figure}
\begin{centering}
\includegraphics[width=0.8\columnwidth]{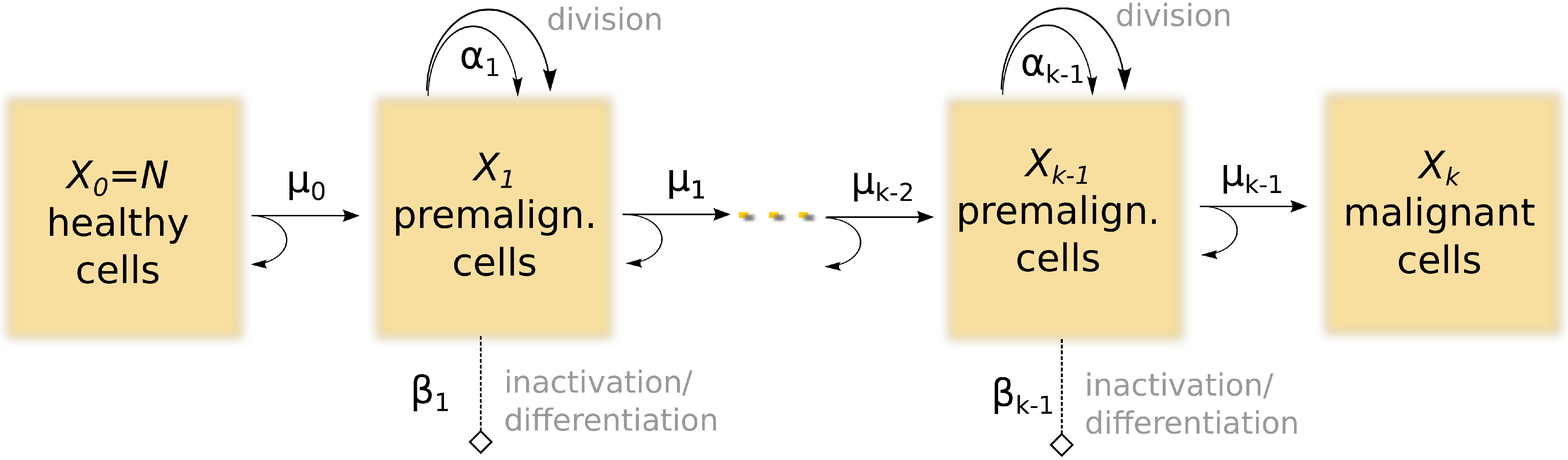}
\par\end{centering}

\begin{centering}
\includegraphics[width=0.53\columnwidth]{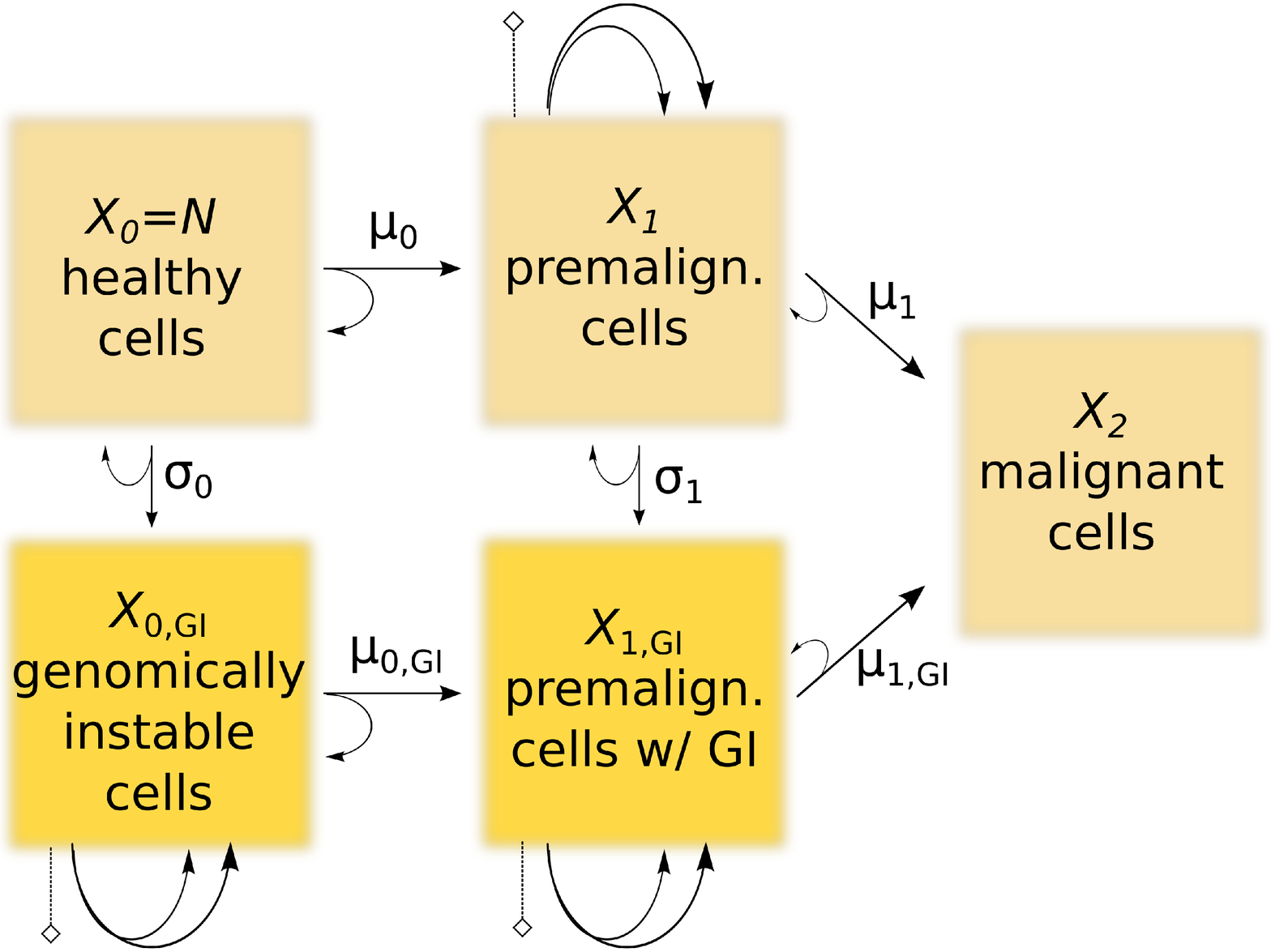}
\par\end{centering}

\protect\caption{\selectlanguage{american}%
Schematic structure of a $k$-stage model (top) and the two-path model
studied in this paper (below). Here, $X_{j}$ denotes the stochastic
number of cells at stage $j$, with arrows indicating transitions
at rates $\mu_{j}$, etc. (see text). Cancer is assumed to occur once
the first malignant cell appears, with latency period $t_{\mathrm{lag}}\sim5$
years. \label{fig:cartoon}\selectlanguage{english}%
}
\end{figure}
\foreignlanguage{american}{Mathematically, this is modeled as continuous-time
Markov processes for the (stochastic) numbers of cells, $X_{i}(t)$,
at the different stages ($i=0,\dots,k$). Specifically, at age $t=0$,
one starts with $X_{0}\equiv N$ healthy cells, which can make a transition
(modeled as a Poisson process) with rate $\mu_{0}$ to a first, premalignant
stage, $X_{1}$.}%
\footnote{\selectlanguage{american}%
Note that the precise genetic or histopathological meaning of these
(pre-)malignant stages is far from clear for all but a few cancer
types. Their number can be very large, and the stages may even differ
strongly between different histological subtypes. The rates pertaining
to these stages are thus considered as giving effective time scales,
determined from fits to the data.\selectlanguage{english}%
}\foreignlanguage{american}{ These cells can then undergo a birth/death
process with rates $\alpha_{1}$/$\beta_{1}$ , leading to a net proliferation
rate $\gamma_{1}\approx\alpha_{1}-\beta_{1}$. Further transitions
eventually lead to malignant cells, $X_{k}$. The occurrence of the
first malignant cell is assumed to lead to cancer after an effective
lag time $t_{\mathrm{lag}}$, typically on the order of a few years.
A cartoon depiction is shown in Fig.~\ref{fig:cartoon}; for $k=2$,
this corresponds to the standard two-stage model with clonal expansion
(TSCE). }

\selectlanguage{american}%
The mathematical model above can be solved for the survival function
$S(t)$ and, equivalently, the hazard (here: lung-cancer mortality
rate), $h=-\frac{d}{dt}\ln S$, using the method of characteristics
\cite{Moolgavkar1988}. For the two-stage model, assuming rate parameters
to be piecewise age-independent, an exact closed-form solution can
be attained \cite{Heidenreich1997}. In the general case, an extension
of this solution is valid approximately if age bins are small enough
for intermediate-cell numbers to change slowly.

As an illustration, let us highlight some generic features shared
by all such multistage models. At earlier ages, the hazard is well
described by a deterministic model, $h\simeq\mu_{k-1}\bar{X}_{k-1}$,
in terms of the mean numbers of cells, $\dot{\bar{X}}_{j}=\mu_{j-1}\bar{X}_{j-1}+\gamma_{j}\bar{X}_{j}$
\cite{Moolgavkar1988}. This leads to an initially polynomial growth,
$h(t)\simeq N\mu_{0}\cdots\mu_{k-1}t^{k-1}/(k-1)!$, followed by a
rapid proliferation-driven phase, $h(t)=\mathcal{O}(e^{\gamma t})$,
where $\gamma$ denotes the maximum growth rate. However, this deterministic
approximation fails to account for the effective reduction in available
premalignant cells as new malignancies arise: Whenever a person reaches
the cancer endpoint, those cells can no longer lead to further cancer
cases. At older ages, approximately around the mean cancer age, a
steady state is reached between growth and effective ``loss'' of
premalignant cells \cite{Moolgavkar1988}, and the hazard levels off
to a constant limit, $h_{\infty}\sim N\mu_{0}\gamma_{1}/\alpha_{1}$.
These borderline cases also illustrate a more general point: Not all
biological rates can be determined from fits to the cancer data alone.
Generically, only certain parameter combinations are identifiable
\cite{heidenreich1996parameters}. 

In the discussion so far, we have tacitly assumed a linear series
of transitions for simplicity. However, we also test a model where
mutations may occur along two different pathways, as sketched in Fig.~\ref{fig:cartoon}(bottom).
Such a model has been introduced by Little \emph{et al.} \cite{Little2003}
to account for genomic instability, inspired by models for colon cancer
\cite{Nowak2002}. The underlying idea is that a second path, activated
via transition rates $\sigma_{j}$, corresponds to the loss of a gene
involved in maintaining genomic integrity. This may lead to mutation
rates much larger than for the genomically stable (upper) path, $\mu_{j}^{\mathrm{GI}}\gg\mu_{j}$
\cite{Eidemuller2014}. Despite the more complex topology, the model
equations are constructed and solved using exactly the same principles
as outlined above.

\subsubsection{Risk modeling}

In the framework of multi-stage models, the hazard is fully determined
by the (generally time-dependent) mutation and growth rates. These
parameters are essentially assumed to have an age-independent background
value which may be modified by external risk factors such as radiation
dose rate, $d(t)$, and smoking, $s(t)$, here assumed to start from
18 years on. In practice, both risk factors are allowed to independently
increase any of the rate parameters $\vartheta_{l}\in\{\mu_{0},\mu_{1},\dots,\gamma_{1},\dots\}$;
from all possible combinations $\{\vartheta_{l}(d),\vartheta_{l'}(s)\}$,
the best-fitting model is selected. Depending on which of these rates
show a radiation effect, the radiation risk varies in a characteristic
fashion with age \cite{Heidenreich1997}, as well as with modifiers
such as duration of or age at exposure.

This is in marked contrast to conventional descriptive models employed
in radiation epidemiology \cite{Preston2003}, which we also use to
benchmark our results. In a descriptive model, the hazard function
is modeled directly -- rather than its underlying mechanism. Here
we use the conventional parametrization \cite{Gilbert2013} 
\begin{equation}
h(t)=h_{\mathrm{bsl}}(t)\left[1+\mathrm{ERR}(D,t,\dots)\right],\label{eq:ERR-model}
\end{equation}
where the baseline, $h_{\mathrm{bsl}}(t)=e^{\psi(t)+\psi_{\mathrm{rf}}}$,
is factored into terms for age-dependent background, $\psi(t)\equiv\sum_{j=0}^{2}c_{j}\ln^{j}(\frac{t}{60\mathrm{a}})$,
and other risk factors, $\psi_{\mathrm{rf}}$ (birth year, smoking
and alcohol, etc.). The excess relative risk, ERR, is factored into
dose-response shape---typically a function of cumulative dose, $D(t-t_{\mathrm{lag}})$---and
time-dependent modifiers such as attained age or age at exposure,
see Appendix.

\subsubsection{Fit procedure}

All model parameters are estimated by maximizing their likelihood
$L$, which is constructed using the individual likelihoods of all
cohort members \cite{mccullagh1989}. Equivalently, we minimize the
deviance, $\mathcal{D}=-2\ln L>0$, using the Minuit function-minimization
library \cite{james1975minuit}. For model selection, we rely on the
likelihood-ratio test for nested models so as to retain only parameters
significant at a 95\% confidence level. The same level is also adopted
throughout this paper for confidence intervals. To rank non-nested
models, the entropy-based Akaike index is used \cite{Akaike1974},
$\mathrm{AIC}=\mathcal{D}+2n$, which effectively penalizes overfitting
for models with larger number of parameters $n$.

\section{Results}

\begin{flushleft}
\begin{table}
\raggedright{}\protect\caption{Synopsis of the best models in this study, along with figures of merit
for their goodness of fit (see text for details). The columns labeled
$d$ and $s$ indicate the model's parametric dependence on dose(rate)
and smoking-related confounders; e.g., $\gamma=\gamma^{(0)}(s)+\delta\gamma(d)$
for the TSCE model. \label{tab:model-fits}}
\begin{tabular}{cccccc}
\hline 
Model & $s$ & $d$ & \# Parameters%
\footnote{Counting parameters for background, smoking/alcohol, birthyear , and
Pu; see Appendix. %
} & Deviance & AIC\tabularnewline
\hline 
descriptive & $\psi_{cf}$ & $\mbox{\textrm{ERR}}(D)$ & 9 & $4770.3$ & $4788.3$\tabularnewline
TSCE & $\gamma$ & $\gamma$ & 10 & $4760.7$ & $4780.7$\tabularnewline
3SCE(A) & $\mu_{1}$ & $\gamma_{1}$ & 11 & $4749.9$ & $4771.9$\tabularnewline
3SCE(B) & $\gamma_{2}$ & $\gamma_{1}$ & 11 & $4756.7$ & $4778.7$\tabularnewline
3SCE(C) & $\gamma_{1}$ & $\gamma_{2}$ & 11 & $4757.7$ & $4779.7$\tabularnewline
\hline 
\end{tabular}
\end{table}

\par\end{flushleft}

Table~\ref{tab:model-fits} provides an overview of the best-fitting
models. Before explaining in detail their mechanisms and the implications
for radiation risk, let us anticipate some general patterns. All highest-ranked
multi-stage models share a Plutonium-induced enhancement of proliferation
rates. More specifically, the fits suggest that 3-stage models with
a radiation effect on an early stage of proliferation (models A, B)
may yield an improved description of the Mayak data, compared with
an effect on the penultimate stage (C). Although model A exhibits
by far the lowest AIC, we will present the radiation risks for several
three-stage models so as to give an impression of the range of model
predictions, as discussed in the Appendix. 

No evidence is found for a radiation-induced second pathway. We note
that all dose rates here refer to internal Pu exposure; external radiation
is not significant in two-stage and descriptive models and thus not
considered in the following.

\subsection{Two-stage model}

The two-stage model has been applied to previous follow-ups of the
Mayak workers \cite{Jacob2005,Jacob2007}. We shall therefore discuss
it here as a benchmark, but also to illustrate some mechanisms inherent
in any multistage model.

\selectlanguage{english}%
\begin{figure}
\begin{centering}
\includegraphics[width=0.7\columnwidth]{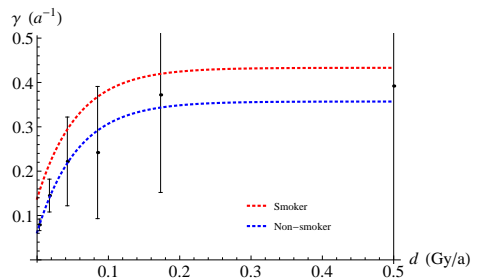}
\par\end{centering}

\protect\caption{\selectlanguage{american}%
Dependence of the proliferation rate, $\gamma(d)$, on internal (lung)
dose rate in the TSCE model. For comparison, the dots illustrate the
results of a categorical fit (with 95\%-level errors). \label{fig:gamma-doserate}\selectlanguage{english}%
}
\end{figure}
\foreignlanguage{american}{We find that the effects of both Pu dose
rate as well as smoking status  are by far best described as additively
enhancing the net proliferation rate, $\gamma(s,d)=\gamma^{(0)}(s)+\delta\gamma(d)$.}%
\footnote{\selectlanguage{american}%
Note that an additional radiation effect on the initiating rate, $\mu_{0}(d)$,
is highly insignificant when fitted to the data and thus not included
in our model. This may be simply due to lack of data power rather
and does not imply irrelevance of radiation-induced initiating events.\selectlanguage{english}%
}\foreignlanguage{american}{ A categorical fit of the dose dependence
strongly suggests a saturation at larger dose rates (Fig.~\ref{fig:gamma-doserate}),
and we find it best modeled by an exponentially leveling function
\begin{equation}
\delta\gamma(d)=\gamma_{\infty}\left(1-e^{-r\times d/\gamma_{\infty}}\right).\label{eq:gamma-dose}
\end{equation}
 Here $r\sim5/\mathrm{Gy}$ (see Table~\ref{tab:model-parameters})
governs the linear, low-dose response, $\delta\gamma\sim r\, d$,
and $\gamma_{\infty}\sim0.3/\mathrm{a}$ denotes the rate approached
as $d\gg d_{*}\equiv\gamma_{\infty}/r$; here $d_{*}\sim0.06\mathrm{Gy/a}$.
This is qualitatively in line with previous analyses \cite{Jacob2005,Jacob2007}
but also several Radon-risk studies \cite{Luebeck1999,hazelton2001analysis,Heidenreich2004,Heidenreich2012,Eidemuller2012},
see Discussion. It is worth noting that the data fit equally well
to a response in terms of the accumulated dose $D$ -- i.e., $\delta\gamma(D)$
in (\ref{eq:gamma-dose}) -- which may relate to the long protracted
exposures of Mayak workers. }

\selectlanguage{american}%
\begin{figure}
\subfloat[Hazard $h(t)$ of the two-stage model for non-/smokers. Shown are
values without (baseline) and with exposure. (Smoking is assumed to
start at age $18$.) \label{fig:Scenario-hazard}]{\includegraphics[width=0.8\columnwidth]{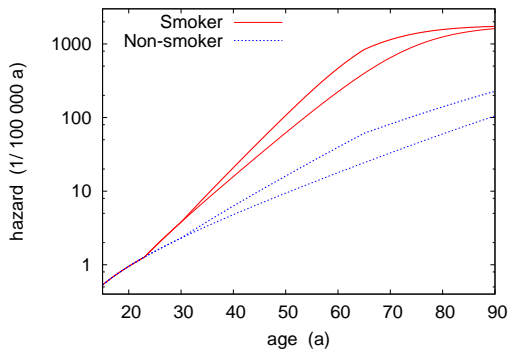}

}

\subfloat[Excess-relative-risk (ERR) ratio of smokers and non-smokers, shown
for different multi-stage models (see text). A ratio of $1$ indicates
multiplicative Plutonium-smoking interaction, as in the standard descriptive
model. The thin-dotted line denotes the (non-significant) sub-multiplicative
trend suggested by an extended descriptive model. \label{fig:ERR_smk-vs-nonsmk}]{\includegraphics[width=0.8\columnwidth]{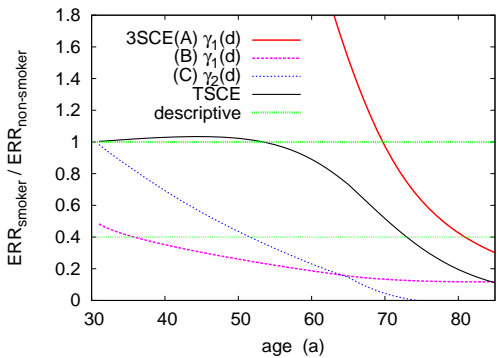}

}

\protect\caption{Risk for a scenario with constant exposure between ages 25 and 60,
at $D=0.2\mathrm{Gy}$. Notice the lag time of $5\mathrm{a}$.\label{fig:Pu-smk-interaction}}
\end{figure}

It should be stressed that, even though the main risk factors, smoking
and Plutonium, enter the growth rate additively, the actual risk will
exhibit an interaction between them. To illustrate this, in Fig.~\ref{fig:Scenario-hazard}
we display the age-dependent hazard, $h(t)$, for a representative
exposure scenario at a constant dose rate from age $t_{1}=25\mathrm{a}$
to $60\mathrm{a}$, with dose $D=0.2\mathrm{Gy}$. For a wide age
range, coinciding with the phase of exponential proliferation, the
relative risks of radiation and smoking approximately multiply (note
the log scale). It is only at larger ages ($\gtrsim60\mathrm{a}$)
that the combined risk drops below the multiplicative value. Such
sub-multiplicity agrees with trends glimpsed in a descriptive analysis
\cite{Gilbert2013}. Here, it follows naturally because the hazard
of Pu-exposed smokers levels off much earlier, reflecting an earlier
onset of cancer (see Methods). 

To better link these findings to those of descriptive models (see
Appendix), we will from now on consider the excess relative risk,
$\mathrm{ERR}\equiv h/h_{\mathrm{bsl}}-1$, defined relative to the
zero-exposure baseline risk. From this angle, multiplicity of risk
implies an equal ERR for smokers and non-smokers -- i.e., a ratio
$\mathrm{ERR}^{(s=1)}/\mathrm{ERR}^{(s=0)}=1$. Figure~\ref{fig:ERR_smk-vs-nonsmk}
shows that, under the scenario above, the risk is indeed multiplicative
until older ages ($t\lesssim60\mathrm{a}$). It then becomes sub-multiplicative,
and the ERR ratio drops markedly below unity after (time-lagged) exposure
ends.

\begin{figure}
\includegraphics[width=0.8\columnwidth]{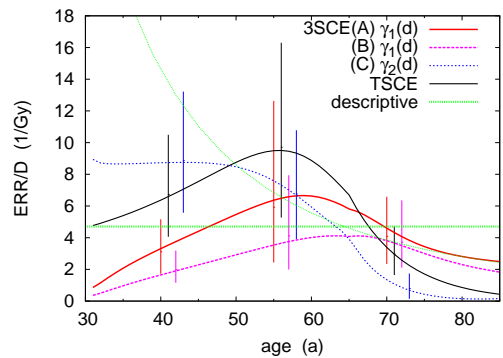}

\protect\caption{Age-dependent excess relative risk ($\mathrm{ERR}/D$) of different
multi-stage models, for smokers with constant exposure between ages
25 and 60 ($D=0.5\mathrm{Gy}$). For comparison, the non-significant
trend in the descriptive model (thin-dotted line) is also shown. (All
error bars are at 95\% confidence level.) \label{fig:Scenario-age}}
\end{figure}
Since most cases are related to smoking, we will now concentrate on
the risk for smokers. Moreover, to separate age and dose dependencies,
we scale the ERR by the accumulated dose, $\mathrm{ERR}(D;t)/D(t-t_{\mathrm{lag}})$,
with the lag time $t_{\mathrm{lag}}=5\mathrm{a}$.  Figure~\ref{fig:Scenario-age}
displays the age-dependent $\mathrm{ERR}/D$ for the scenario above
but at $D=0.5\mathrm{Gy}$. For the two-stage model, it reveals a
characteristic increase during exposure (owing to proliferation),
followed by a marked drop-off after exposure has ended. The latter
is similar to the (non-significant) attained-age trend found in the
descriptive model, cf. Appendix.

\begin{figure}
\subfloat[$\mathrm{ERR}/D$ for different ages at first exposure, $t_{1}$ (at
fixed duration $\Delta t=20\mathrm{a}$). \label{fig:Scenario-aae}]{\includegraphics[width=0.8\columnwidth]{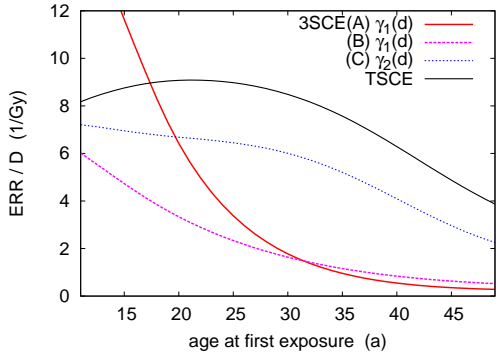}

}

\subfloat[$\mathrm{ERR}/D$ for different durations $\Delta t$ (at fixed $t_{1}=20\mathrm{a}$).
\label{fig:Scenario-dur}]{\includegraphics[width=0.8\columnwidth]{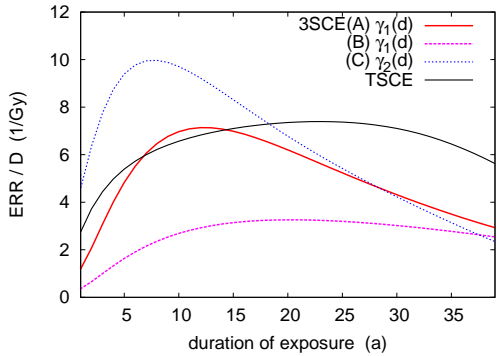}

}

\protect\caption{Dependence of $\mathrm{ERR}/D$ on different exposure patterns, for
smokers with $D=0.2\mathrm{Gy}$. \label{fig:Scenario-dependence}}
\end{figure}
The age dependence just discussed pertains to one specific scenario.
Let us now indicate how the risk is modified by different exposure
patterns. The dependence upon age at first exposure, $t_{1}$, is
shown in Fig.~\ref{fig:Scenario-aae}. We have chosen a scenario
with $D=0.2\mathrm{Gy}$ and a typical duration $\Delta t=20\mathrm{a}$;
the ERR is recorded at $t_{1}+\Delta t+t_{lag}$. Clearly, for this
two-stage model, the variation is rather mild for all but very early
exposures (not encountered at Mayak) and very late ones. In the former
case, virtually no premalignant cells are available for proliferation.
At older ages, in turn, they are increasingly lost to new malignancies;
thus the risk is attenuated. In the descriptive model, a weak (non-significant)
trend also suggests a slight decrease of ERR with older ages at exposure
(not shown). 

Figure~\ref{fig:Scenario-dur} reveals a characteristic influence
of exposure duration, $\Delta t$ (where $D=0.2\mathrm{Gy}$, $t_{1}=20\mathrm{a}$).
For very short exposures, $\Delta t\ll D/d_{*}\sim3\mathrm{a}$, the
ERR is strongly suppressed: The short duration cannot be compensated
by an increased dose rate, since the growth rate saturates at dose
rates larger than $d_{*}$. This inverse dose-rate effect is thus
inherently connected to a saturating radiation response as in Eq.~(\ref{eq:gamma-dose}).
It has been observed also in mechanistic Radon studies \cite{Luebeck1999,hazelton2001analysis,Eidemuller2012},
although the Mayak data are not powerful enough to support this at
the descriptive level. At sufficiently long exposures, $\Delta t\gg D/d_{*}$,
the inverse dose-rate effect disappears and eventually gives way to
a slight direct effect. This is related to the leveling of the hazard
upon reaching malignancy. 

\begin{figure}
\includegraphics[width=0.8\columnwidth]{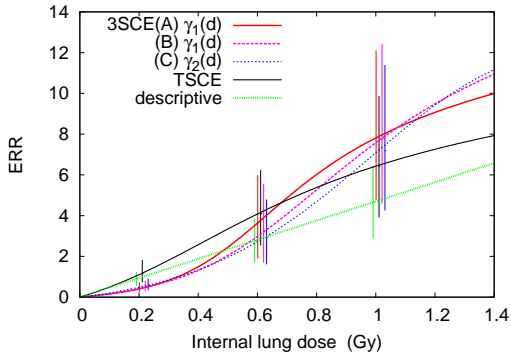}

\protect\caption{Dose-response relationship, $\mathrm{ERR}(D)$, for exposure during
ages 25 to 60, recorded at age $t=65$. For comparison, the linear
response $\mathrm{ERR}=cD$ is plotted. \label{fig:dose-response}}
\end{figure}
To wrap up this discussion, let us consider the dose-response relationship
implied by this proliferation-based model. Figure~\ref{fig:dose-response}
illustrates that, in contrast to the linear dose response typically
assumed in descriptive modeling ($\mathrm{ERR}(D)=cD$, here $c\approx4.7/\mathrm{Gy}$),
this model exhibits a nonlinear response. Although the quantitative
values depend on the exposure scenario (here: exposure during $t=25-60\mathrm{a}$),
some general features apply to any two-stage model with a proliferation
enhancement similar to Eq. (\ref{eq:gamma-dose}): The dose response
is characterized by a linear low-dose regime, $\mathrm{ERR}\simeq r\times D$
for $D\ll1/r$, and leveling at sufficiently large dose(rate)s and/or
ages. Intermittently, typically an exponential increase is seen, reflecting
the exponential growth of premalignant cells due to proliferation.
However, this effect tends to be washed out by the leveling at older
ages. In a descriptive model fit, where essentially an averaging occurs
over all exposure histories in the cohort, this pronounced nonlinearity
will be even harder to resolve. Still, it is noteworthy that a TSCE-inspired
descriptive model, $\mathrm{ERR}=e^{f(D)}-1$ with $f(D)\equiv c_{\infty}(1-e^{-c_{\mathrm{lin}}D/c_{\infty}})$,
yields an improved fit (but similar AIC), with a linear response $c_{\mathrm{lin}}\sim3.2/\mathrm{Gy}$
and leveling at $D_{*}\equiv c_{\infty}/c_{\mathrm{lin}}\sim0.8\mathrm{Gy}$.

\subsection{Three-stage models}

As mentioned earlier, the highest-ranked 3-stage models fall into
two categories. 

The best two models (A, B; see Table~\ref{tab:model-parameters})
show a radiation effect on the \emph{earlier }stage of proliferation,
$\gamma_{1}(d)$, differing only in their background parameters --
specifically the smoking response on $\mu_{1}$ (A) or $\gamma_{2}$
(B). This early impact leads to a substantially delayed radiation
response compared to both the 2-stage model and 3-stage model C, as
is seen from Fig.~\ref{fig:Scenario-age}: The $\mathrm{ERR}/D$
is initially zero and, once it has peaked, tends to drop off more
mildly. (In some scenarios, the $\mathrm{ERR}/D$ may even increase
when exposure stops.) This lag is intuitive, as the insulted cells
need to pass through an additional mutation stage before becoming
malignant. In stark contrast, model C---with an effect on the penultimate
stage---predicts a much higher risk, $\mathrm{ERR}/D\simeq r\sim10/\mathrm{Gy}$,
right after exposure. In the scenario depicted in Fig.~\ref{fig:Scenario-age},
it further displays an almost monotonic drop with attained age, not
unlike the trend seen in the descriptive model risk. 

However, it must be stated that this specific scenario conceals an
underlying complexity not present in the 2-stage model. The reason
is that the 3-stage model is determined by two competing growth rates,
$\gamma_{1}$ and $\gamma_{2}$. Let us consider models A/B: For a
small enough dose rate such that $\gamma_{1}^{(0)}+rd<\gamma_{2}^{(0)}$,
the clonal dynamics is governed by the largest baseline rate, $\gamma_{2}^{(0)}$.
In other words, below the critical dose rate $d_{\mathrm{crit}}\equiv(\gamma_{2}^{(0)}-\gamma_{1}^{(0)})/r$$\sim0.01\mathrm{Gy/a}$,
both baseline and excess risk grow with the same exponential rate
-- hence the ERR would level off with age. It is only above that critical
dose rate that an exponential increase is seen in the ERR. Likewise,
for model C, the critical point, $(\gamma_{1}^{(0)}-\gamma_{2}^{(0)})/r$,
marks the dividing line between exponential increase and leveling
with age. Notice that the value in Fig.~\ref{fig:Scenario-age} is
just at the borderline.

The dose-rate dependence just described is reflected in the dose response
(Fig.~\ref{fig:dose-response}). At low doses, a linear increase
is seen, just as for the two-stage model. However, that low-dose response
is typically much lower than what we saw for two-stage and descriptive
models. It is only at the critical dose, here $d_{\mathrm{crit}}\Delta t\sim0.35\mathrm{Gy}$,
that a rapid exponential increase sets in. Compared to the two-stage
case, this exponential increase is much sharper owing to the higher
response coefficients, $r\sim(10-17)/\mathrm{Gy}$. As before, the
response levels off once the corresponding dose rates exceed $d_{*}\sim(0.03-0.05)\mathrm{Gy/a}$. 

Another consequence of an early-stage radiation effect is a notably
suppressed risk for older ages at exposure, $t_{1}$ (Fig.~\ref{fig:Scenario-aae}).
To understand this, recall that the radiation effect consists in multiplying
the available pool of stage-1 cells, $X_{1}(t_{1})$, prior to their
making the transition to stage 2. Since that number grows only very
slowly at a rate $\gamma_{1}^{(0)}<\gamma_{2}^{(0)}$, the head start
given to already existing stage-2 cells, $X_{2}(t_{1})$, becomes
overwhelming the later in life exposure starts. Thus the ERR is suppressed
exponentially as $e^{-(\gamma_{2}^{(0)}-\gamma_{1}^{(0)})t_{1}}$
for large $t_{1}$. By contrast, the mechanism for model C is fairly
similar to that of the 2-stage model: Both radiation and baseline
risk are governed by the growth of existing stage-2 cells. This is
why virtually no dependence on $t_{1}$ is seen other than a mild
drop-off for older ages at exposure, due to the onset of malignancies.

Differences with respect to the two-stage model may also be observed
in the exposure-duration dependence (Fig.~\ref{fig:Scenario-dur}).
Like the 2-stage models, all 3-stage models exhibit an inverse dose-rate
effect for durations $\Delta t\lesssim D/d_{*}$. As explained previously,
this merely hinges on the saturation of the proliferation rate for
higher dose rates. It may be noted that the early-stage models A/B
lead to a stronger risk suppression at such short durations, due to
the extra mutational step to be passed. However, a more qualitative
difference is found for longer durations: Here a marked direct effect
occurs, the ERR falling off as $\Delta t{}^{-1}$. This may be viewed
as a linear dose-rate modification for perturbatively small dose rates
$D/\Delta t$ in the limit of long durations.

Finally, let us comment on the interaction between radiation and smoking
risks. We saw earlier that the two-stage model predicts a largely
multiplicative interaction or, equivalently, a near-unit ratio of
smokers' and non-smokers' ERR (Fig.~\ref{fig:ERR_smk-vs-nonsmk}).
The situation is less clear cut for the three-stage models. Model~C,
with a later-stage radiation effect, is most comparable to the 2-stage
case: Initially, it also exhibits near-multiplicity, which reflects
that radiation simply leads to multiplication of existing stage-2
cells. However, the ERR ratio then drops very rapidly. This is essentially
because for smokers, the dose rate is \emph{below }the critical value,
$(\gamma_{1}^{(0)}-\gamma_{2}^{(0)})/r\sim0.01\mathrm{Gy/a}$, as
smoking leads to a very large growth rate $\gamma_{1}(s=1)\approx0.16\mathrm{a^{-1}}$.
Thus the ERR levels off, in contrast to the exponentially growing
ERR for non-smokers, where there is no threshold. An analogous mechanism
is at work in the model B, which is sub-multiplicative throughout. 

By contrast, for model A, the risk in the scenario shown in Fig.~\ref{fig:ERR_smk-vs-nonsmk}
is strikingly super-multiplicative except for ages $t\gtrsim70\mathrm{a}$.
At first glance, this deviation from multiplicity might seem surprising:
For constant rates, the hazard is typically proportional to $\mu_{1}(s)$,
and the smoking dependence should thus drop out of the relative risk.
However, smoking is assumed to start at age 18, only a few years prior
to irradiation. Hence the baseline risk---initially proportional to
the number of existing stage-2 cells---mostly stems from those cells
created \emph{before }smoking started, and is thus comparable for
non-/smokers. Thus, the large ERR for smokers reflects the much higher
excess risk due to freshly mutated stage-1 cells. It is only long
after the start of exposure that the smokers' baseline risk increases
sufficiently to compensate for this.

\subsection{Two-pathway models }

We have tested a family of multi-path models of genomic instability
(GI). Let us briefly outline the key assumptions made here to reduce
the number of parameters -- in total, 6 mutation rates and 6 cell-division/death
rates, as sketched in Fig.~\ref{fig:cartoon}. Most premises are
motivated by the biological mechanisms thought to underlie GI \cite{Nowak2002,Eidemuller2014}.
First, the destabilizing transition rates are set equal, $\sigma_{j}\equiv\sigma$,
as they pertain to the inactivation of the same genomic-integrity
gene. Since, presumably, GI \emph{per se} does not lead to any growth
advantage, we set the birth/death rates at stage $0^{\mathrm{GI}}$
equal, $\alpha_{0}^{\mathrm{GI}}=\beta_{0}^{\mathrm{GI}}$, their
precise value being marginal. Mutation rates following GI, $\mu_{j}^{\mathrm{GI}}$,
are supposed to be larger, or at least equal, compared to their genomically
stable (upper-branch) counterparts.

Although GI may well be present as a sporadic mechanism, the baseline
data are not thought to provide sufficient structure to distinguish
complex background models (see Discussion; cf. also Ref.~\cite{Little2007}).
We have thus focused on the case of radiation-induced GI -- i.e.,
an activation of the otherwise silent GI path ($\sigma$) by radiation.
This may be modeled as $\sigma=r_{\mathrm{GI}}\times d$ (with vanishing
background rate). In addition, radiation may affect regular mutation
or growth rates.

None of the tested two-pathway models has led to any significant,
numerically stable improvement beyond the benchmark two-stage model.
It is stressed that relaxing the assumptions of linearity or vanishing
background rate do not yield a significant improvement. 

\begin{table*}
\begin{raggedright}
\protect\caption{Observed numbers of lung-cancer cases by dose categories, compared
with those predicted by the descriptive and multi-stage models (in
brackets: excess cases). As a reference, we also give the person years
(py).\label{tab:dose-statistics}}
\begin{tabular}{c|c|cccccc}
\hline 
Doses (Gy) & py & Cases$\;$ & TSCE$\;$ & 3SCE(A)$\;$ & 3SCE(B)$\;$ & 3SCE(C)$\;$ & descriptive\tabularnewline
\hline 
$0-0.01$ & $188,995$ & $204$ & $198$ ($0.5$) & $203$~($0.2$) & $205$~($0.2$) & $204$~($0.3$) & $194$~($0.6$)\tabularnewline
$0.01-0.03$ & $18,888$ & $32$ & $50$ ($3$) & $49$~($1$) & $50$~($1$) & $50$~($1$) & $49$~($4$)\tabularnewline
$0.03-0.1$ & $15,843$ & $58$ & $51$ ($9$) & $47$~($4$) & $47$~($4$) & $48$~($4$) & $52$~($11$)\tabularnewline
$0.1-0.3$ & $7,091$ & $41$ & $37$ ($15$) & $31$ ~($8$) & $30$~($23$) & $31$~($7$) & $39$~($17$)\tabularnewline
$0.3-1$ & $3,153$ & $29$ & $33$ ($24$) & $34$~($25$) & $31$~($21$) & $31$~($21$) & $32$~($23$)\tabularnewline
$>1$ & $738$ & $24$ & $18$ ($16$) & $24$ ~($21$) & $25$ ~($22$) & $25$~($22$) & $22$~($20$)\tabularnewline
\hline 
total & $234,708$ & $388$ & $388$ ($67$) & $388$~($59$) & $388$~($56$) & $388$~($56$) & $388$~($75$)\tabularnewline
\hline 
\end{tabular}
\par\end{raggedright}

\end{table*}

\section{Discussion}

\subsection{Mechanism of radiation action}

A robust result of our analysis is an $\alpha$-radiation-induced
enhancement of proliferation rates of premalignant cells. This corroborates
a consistent finding in many studies based on fitting two-stage models
to lung-tumor data, both from epidemiological cohorts (Radon-exposed
miners \cite{Luebeck1999,hazelton2001analysis,Heidenreich2004,Heidenreich2012,Eidemuller2012})
and animal experiments (see Ref.~\cite{heidenreich2000analysis}
for an overview). It also alleviates concerns that a proliferation
effect might be confined to the two-stage model \cite{Little2002}. 

However,  it has long been criticized \cite{brugmans2002overrated}
that no radiobiological evidence exists for such a premalignant-growth-enhancing
effect.  Although experimental evidence is still sparse \cite{Kopp-Schneider2006,chao2008preneoplastic},
let us briefly discuss some theoretical models put forward to explain
how radiation might lead to enhanced proliferation of premalignant
cells.

The most elaborate model is based on the following idea \cite{Heidenreich2001}:
Radiation kills cells, which in turn triggers division of neighboring
stem cells so as to replenish the lost tissue. This may lead to net
proliferation of premalignant cells if these have a selective advantage,
that is, a slightly higher rate for cell division than needed for
homeostasis. (Strictly speaking, cells need not necessarily be killed;
it might be sufficient for them to have a proliferative disadvantage---e.g.,
by effected cell-cycle arrest---such that these are subsequently repelled
by premalignant cells.) That hypothesis has been shown \cite{Bijwaard2006,Heidenreich2008,Madas2014}
to lead to a dose-rate response, $\gamma(d)$, which is qualitatively
similar to that found in cohort studies (see, e.g., Fig.~\ref{eq:gamma-dose}),
albeit with quantitatively modest agreement. Saturation of the growth
rate much higher than a characteristic dose rate, $d\gg d_{\mathrm{s}}$,
occurs if more cells are killed than can be substituted for by premalignant
cells. 

To obtain a very rough estimate for that critical dose rate, note
that $\alpha$-particle hits of a cell nucleus are independent and
rare. Thus the number of hits, $N$, is Poisson-distributed, $P_{N}=e^{-\bar{N}}\bar{N}^{N}/N!$,
which means the fraction of cells \emph{not }hit is $P_{N=0}=e^{-\bar{N}}$.
Assuming (i) a linear dose response, $\bar{N}=n\, D$, with $n\sim4/\mathrm{Gy}$
\cite{beir1999health}, and (ii) delivery of the dose $D\equiv d\,\tau$
over a characteristic time of order the interval between cell cycles
($\tau\sim1\mathrm{a}$ for basal stem cells \cite{harley1996biological}),
we have $P_{N=0}(d)=e^{-nd\tau}$. At the characteristic dose rate,
$d_{\mathrm{s}}$, about one out of, say, six nearest neighbors would
be hit, such that $P_{0}(d_{s})=1-p$, $p\equiv1/6$, yielding a characteristic
dose rate of 
\begin{equation}
d_{s}\simeq\frac{p}{n\tau}\sim0.04\frac{\mathrm{Gy}}{\mathrm{a}}.\label{eq:Dose-leveling}
\end{equation}
This is on the same order of magnitude as the value found in this
study, $d_{*}\sim(0.03-0.06)\mathrm{Gy/a}$. In this light, the model
results presented here and the repopulation mechanism may be interpreted
to be compatible. However, it is important bearing in mind that these
estimates are naturally crude. Matters are further complicated by
the spatially inhomogeneous energy deposition within different spots
in the lungs, an effect not reflected in whole-lung doses used here
\cite{Balashazy2009}.

As an alternative mechanism, a radiation-induced disturbance of cell
communication has been suggested \cite{Curtis2004,Jacob2010}. This
may lead to, e.g., up-regulated growth signals or a reduction of apoptosis
\cite{Trosko2005,Kundrat2014}, with a higher effect on intermediate
cells because those tend to evade homeostatic control. It has even
been proposed that a proliferation enhancement, mediated by such a
bystander signaling, might be the generic mechanism for the response
to densely ionizing radiation \cite{Shuryak2011}. However, no mechanistic
model has been put forward explaining in detail how this might lead
to a dose-rate response, $\gamma(d)$. 

Even so, should the radiation response indeed be governed by the bystander
effect, then a similar behavior ought to be expected as for bystander-mediated
mutation induction \cite{Brenner2001}.  In microbeam experiments
\cite{sawant2001bystander,Zhou2001}, it has been found that for low
doses---corresponding to less than $\sim10\%$ of cells being hit---the
mutagenic yield was strongly amplified as bystander cells also received
signals. (A similar pattern has been found for intercellular induction
of apoptosis \cite{abdelrazzak2011role}.) For much higher doses,
in turn, the response essentially saturated. Along the lines leading
to Eq.~(\ref{eq:Dose-leveling}), we can estimate the characteristic
dose rate $d_{s}$ by assuming the crossover to occur at a fraction
$p=0.1$ of cells being hit. This yields $d_{s}\approx p/n\tau\sim0.025\mathrm{Gy/a}$,
under the same caveats as mentioned above. From this standpoint, both
bystander signaling and the repopulation hypothesis appear compatible
with the dose-rate response found in this study.

\subsection{Comparison with previous studies}

As discussed earlier on, for mechanistic models, the risk is essentially
determined by its structure, particularly, the radiation response.
A dose-rate dependent proliferation rate, $\gamma(d)$, saturating
for larger dose rates (as in Eq.~\ref{eq:gamma-dose}) is found in
many studies applying the two-stage model to $\alpha$-particle-induced
lung cancer. In particular, this response quantitatively agrees with
that of the preferred two-stage model by Jacob \emph{et al}. for the
Mayak cohort, both for Plutonium and smoking \cite{Jacob2007}. Concordantly,
their risk estimates are similar to those of the present TSCE model
-- such as a cohort-averaged excess risk of $\mathrm{ERR}(t=60\mathrm{a})/D\sim4/\mathrm{Gy}$,
a nonlinear dose dependence $\mathrm{ERR}(D)$ for larger doses, and
sub-multiplicity of smoking and radiation risks.

In a recent descriptive analysis of the Mayak data, Gilbert \emph{et
al}. found a linear dose response, modified significantly only by
a drop-off with attained age (\cite{Gilbert2013}, see also Appendix).
The value at age 60, $\mathrm{ERR}(t=60\mathrm{a})\sim7D/\mathrm{Gy}$,
is somewhat higher than for the cohort average of the mechanistic
models presented here. By contrast, the multi-stage models exhibit
a strongly nonlinear dose dependence especially for higher doses.
Furthermore, they typically display a decrease with attained age only
for large enough ages, most pronounced after the end of exposure.
Initially, an increase with age is seen due to exponential clonal
growth, at least for high enough doses.

In contrast to descriptive models, where an exposure modifier may
not be significant when parametrized explicitly, mechanistic models
implicitly make predictions for the risk dependence on any exposure
scenario. This is exemplified by the age-at-exposure dependence or
the inverse dose-rate effect shown by (at least some) mechanistic
models (Fig.~\ref{fig:Scenario-dependence}). Furthermore, a non-significant
trend in Ref.~\cite{Gilbert2013} indicated a sub-multiplicative
interaction between Plutonium dose and smoking. This is in agreement
with results from the 2-stage model, which further offers a mechanistic
interpretation in terms of exponential cell growth, combined with
earlier malignancies for smokers (see Results).

\subsection{Implications for lung carcinogenesis}

We have shown that several three-stage models give an improved description
of the data compared to one involving two stages. Evidently, this
stochastic inference is based solely on the lung-cancer endpoint and
cannot replace experimental insight into the dynamics of intermediate
stages. Even so, it is worth emphasizing that the Mayak data do not
fully allow to distinguish general mechanisms for carcinogenesis.
Rather, the results here mostly rely on the radiation-associated risk.
In fact, the deviances of the various mechanistic and descriptive
\emph{baseline }models (risk factors age and smoking) do not differ
noticeably -- in contrast to the models including exposure (Table~\ref{tab:model-fits}).

That statement seemingly contradicts the fact that there is only a
fraction $\sim25\%$ of excess cases relative to the baseline (as
can be seen from $\overline{\mathrm{ERR}}\sim5\bar{D}/\mathrm{Gy}$,
the whole-cohort average dose being $\bar{D}\sim0.05\mathrm{Gy}$).
This translates to $60-70$ excess cases (Table~\ref{tab:dose-statistics}),
compared to about $320$ baseline cases. However, truly spontaneous
baseline cases ($30-40$) are outnumbered by smoking-related ones
by a factor of $\sim10$. Moreover, the smoking variable is only binary
(and noisy). This makes it hard to resolve the actual baseline risk
accurately, especially since the multi-stage models do not differ
in their qualitative behavior except for young ages. By contrast,
the models do differ markedly for different irradiation scenarios
as found in the Mayak data.

\section{Conclusion }

We have shown that carcinogenesis models extended to three stages
offer an improved description of the Mayak lung-cancer data, as compared
to the two-stage model. All favored 3- and 2-stage models indicate
a radiation-enhanced proliferation rate of premalignant cells, suggesting
that this is a robust finding not limited to the framework of two
stages. Despite that structural similarity, the models make qualitatively
different predictions for the risk following certain exposure scenarios.
For instance, those models whose radiation impact is on an earlier
stage exhibit a strongly suppressed risk for older ages at exposure.
As opposed to the two-stage case, all three-stage models reveal a
soft-threshold dose(rate) above which the excess risk increases sharply.
Moreover, while an inverse dose-rate effect is predicted by all models,
only those with three stages also display a pronounced direct effect
for longer exposure durations.

One aim of this study has been to elucidate which aspects of carcinogenesis
models are persistent themes or rather model-specific features. Such
a better understanding should facilitate the development of improved
carcinogenesis models, both mechanistic but also descriptive ones.
There is still some way to go toward a more realistic description.
Ultimately, this would involve biological input on rates or premalignant
stages so as to cut the number of undetermined parameters. Closer
at hand, a natural next step may be validating the current models
on other data sets. An extension to more than three stages conceivably
provides a further improvement. Moreover, to get a more accurate description
of the biological mechanisms, it is desirable to develop models specifically
for the different histological cancer subtypes.
\begin{acknowledgments}
S.Z. thanks R. Meckbach, C. Simonetto, P. Kundrat, B.G. Madas, M.
Rosemann, P. Jacob, J.C. Kaiser, and M. Blettner for fruitful discussions.
This work was supported by the European Commission under FP7 project
no. 269553 (EpiRadBio).
\end{acknowledgments}

\section*{}

\appendix

\section{Parameter estimates}

We now present some details on the model results.

The reference descriptive model closely follows Ref.~\cite{Gilbert2013}.
The main risk factor is smoking, which is included simply as a factor
in Eq.~(\ref{eq:ERR-model}), $e^{\psi_{\mathrm{smk}}}=e^{2.3\pm0.3}\approx10$.
The alcohol status further increases the baseline risk by $e^{\psi_{\mathrm{alc}}}=e^{0.6\pm0.1}\approx1.8$.
In addition, the birth year was found to elevate the risk between
years 1915 and 1935 by about $e^{0.4\pm0.1}\approx1.5$. With 3 extra
parameters, this birth-cohort effect is moderately significant ($\Delta\mathcal{D}=-11$).
It might be interpreted as a smoking modifier (as which it yields
only a slightly higher deviance), possibly related to the changed
smoking levels between the wars. The effect is irrelevant for the
radiation risk, and generally no birth-cohort distinction is made
in the scenarios in Figs.~\ref{fig:Pu-smk-interaction}--\ref{fig:dose-response}.
We mention that a calendar-year dependence has been tested but discarded:
It oscillates strongly but shows no conclusive trend, nor does it
influence the dose response.

The dose dependence is modeled as a linear function, $\mathrm{ERR}=cD$,
$c=(4.7\pm0.9)\mathrm{Gy}^{-1}$, with no evidence for threshold or
quadratic terms. Note that the lag time entering the dose, $D\equiv D(t-t_{\mathrm{lag}})$,
does not alter the deviance significantly between $0-10\mathrm{a}$.
We fix it at $t_{\mathrm{lag}}=5\mathrm{a}$, also for all multi-stage
models. No significant time-dependent dose modifiers are found. However,
a trend suggests a decrease of $\mathrm{ERR}/D$ with attained age,
modeled as $(t/60\mathrm{a})^{-2.4\pm2.5}$, and a marginal decrease
with median age at exposure. Even though the reference model (\ref{eq:ERR-model})
implies multiplicative risks of smoking and radiation, a non-significant
trend indicates sub-multiplicity, with an ERR for smokers reduced
by a factor $e^{-0.9(\pm1.5)}\sim0.4$.

The maximum-likelihood parameters of the TSCE model are shown in Table~\ref{tab:model-parameters}.
We have omitted two covariables not relevant for the radiation risk:
alcohol and birth year. Both are included as addends to the growth
rate $\gamma$, consistent with an interpretation as smoking surrogates.
We have also tested an explicit age dependence of the rate parameters,
which indicated a reduction of $\mu_{1}(t)$ or, equivalently, $\gamma(t)$
above age $\sim50\mathrm{a}$. However, these effects were numerically
unstable and have thus been discarded. To convey an impression of
the fit quality, Table~\ref{tab:dose-statistics} shows a juxtaposition
of observed cancer cases and those predicted by the various models
across the dose range.

\begin{table*}
\raggedright{}\protect\caption{Key parameter estimates (including symmetric 95\%-level uncertainties).
\label{tab:model-parameters}}
\subfloat[Two-stage model]{

\begin{tabular}{cc}
\hline 
Parameter & TSCE\tabularnewline
\hline 
$N\mu_{0}\mu_{1}$ ($\mathrm{a}^{-2}$) & $(3.9\pm3.6)\times10^{-7}$\tabularnewline
$\alpha\mu_{1}$ ($\mathrm{a}^{-2})$ & $(3.0\pm2.0)\times10^{-6}$\tabularnewline
$\gamma$ ($\mathrm{a}^{-1}$) & $0.062\pm0.014$ \tabularnewline
$\delta\gamma(s)$ ($\mathrm{a}^{-1}$) & $0.076\pm0.009$\tabularnewline
$r$ $(\mathrm{Gy}^{-1})$ & $5.2\pm1.1$\tabularnewline
$\gamma_{\infty}$ $(\mathrm{a^{-1})}$ & $0.30\pm0.08$\tabularnewline
\hline 
\end{tabular}}\subfloat[Best three-stage models]{

\begin{tabular}{cccc}
\hline 
Parameter & 3SCE(A) & 3SCE(B) & 3SCE(C)\tabularnewline
\hline 
$N\mu_{0}\mu_{1}\mu_{2}$ ($\mathrm{a}^{-3}$) & $(1.3\pm1.6)\times10^{-9}$ & $(2.0\pm2.6)\times10^{-8}$ & $(4.0\pm4.2)\times10^{-8}$\tabularnewline
$\alpha_{1}\mu_{1}\mu_{2}$ ($\mathrm{a}^{-3})$ & $(1.0\pm1.0)\times10^{-8}$ & $(1.2\pm1.7)\times10^{-7}$ & $(4.7\pm3.4)\times10^{-7}$\tabularnewline
$\alpha_{2}\mu_{2}$ ($\mathrm{a}^{-2})$ & $(8.4\pm8.0)\times10^{-6}$ & $(0.7\pm1.1)\times10^{-5}$ & $+0$%
\footnote{Parameter not converged, set to arbitrarily small, positive value
($\sim10^{-10}$).\label{fn:par-not-converged}%
}\tabularnewline
$\mu_{1}\mu_{2}$ ($\mathrm{a}^{-2})$ & $+0$  & $+0$ & $(1.2\pm1.4)\times10^{-6}$\tabularnewline
$\alpha_{1}-\beta_{1}$ ($\mathrm{a}^{-1}$) & $0$ (n.s.)%
\footnote{Not significant; set to zero.%
} & $0.068\pm0.043$ & $0$ (n.s.)\tabularnewline
$\gamma_{2}$ ($\mathrm{a}^{-1}$) & $0.16\pm0.03$  & $0$ (n.s.) & $0.050\pm0.043$\tabularnewline
(smoking) & $\mu_{1}\times e^{3.6\pm0.8}$ & $\gamma_{2}+$$\left(0.14\pm0.03\right)\mathrm{a}^{-1}$ & $\gamma_{1}+$$\left(0.16\pm0.03\right)\mathrm{a}^{-1}$\tabularnewline
$r$ $(\mathrm{Gy}^{-1})$ & $17\pm7$ & $10\pm5$ & $10\pm4$\tabularnewline
$\gamma_{\infty}$ $(\mathrm{a^{-1})}$ & $0.54\pm0.15$ & $0.46\pm0.15$ & $0.39\pm0.16$\tabularnewline
\hline 
\end{tabular}}
\end{table*}

A comment is in order on the parameter estimates of the three-stage
models (Table~\ref{tab:model-parameters}). Some of the (structurally)
identifiable parameter combinations may be practically indeterminable
insofar as they leave the minimum deviance $\mathcal{D}$ virtually
unchanged. In model A, e.g., $\mathcal{D}$ is independent of $\mu_{1}\mu_{2}\to0$,
which has been fixed at an arbitrarily small value. Furthermore, the
best estimate of $\gamma_{1}^{(0)}=\alpha_{1}-\beta_{1}\approx-0.1\mathrm{a^{-1}}$
is not significant and $\gamma_{1}^{(0)}$ thus is set to zero. Worse
yet, for model B (C), the background rate $\gamma_{2}^{(0)}$ ($\gamma_{1}^{(0)}$)
is unstable, tending to unrealistically large negative values. Since
only the smokers' value is stable, $\gamma_{2}^{(0)}+\Delta\gamma_{2}(s=1)\sim0.1\mathrm{a}^{-1}$,
we have set $\gamma_{2}^{(0)}\equiv0$, even at the price of a higher
deviance. It is also for these intricacies that we have opted to present
several of the best three-stage models, rather than relying strictly
on the weight suggested by Akaike's index. This way, a more plausible
impression is given of the uncertainty of model predictions. 

\bibliographystyle{../../bin/plos2009}
\bibliography{/home/sascha/bib/cancermodels}

\end{document}